\documentclass{iau}
\usepackage{graphicx} 

\title[IAUS291.~~Beam modeling of PSR J1906+0746  ] %% short title %%
{PSR J1906+0746: From relativistic spin-precession to beam modeling} %% full title %%

\author[G. Desvignes et al.]  %% short author list %%
{Gregory Desvignes$^1$,
Michael Kramer$^1$,
Isma\"el Cognard$^2$,\\
Laura Kasian$^3$,
Joeri van Leeuwen$^4$,
Ingrid Stairs$^3$\\
  \and Gilles Theureau$^5$
}

\affiliation{$^1$Max-Planck-Institut f\"ur Radioastronomie, Auf dem H\"ugel, 69 D-53121 Bonn, Germany\\ email: {\tt gdesvignes@mpifr-bonn.mpg.de} \\[\affilskip]
$^2$Laboratoire de Physique et Chimie de l'Environnement et de l'Espace, \\3A Avenue de la Recherche Scientifique, 45071 Orl\'eans cedex 2, France  \\[\affilskip]
$^3$Department of Physics and Astronomy, University of British Columbia, \\6224 Agricultural Road, Vancouver, BC V6T 1Z1, Canada\\[\affilskip]
$^4$ASTRON, the Netherlands Institute for Radio Astronomy, Postbus 2, 7990 AA Dwingeloo, The Netherlands\\
[\affilskip]
$^5$Station de radioastronomie de Nan\c cay, Observatoire de Paris, CNRS/INSU, \\ 18330 Nan\c cay, France 
}

\pubyear{2012}
\volume{291}  %% insert here IAU Symposium No.
\jname{\mbox{Neutron Stars and Pulsars: Challenges and Opportunities after 80 years}}
\editors{J. van Leeuwen, ed.} 
\begin{document}

\maketitle

%% -- Abstract ----------------------------------
\begin{abstract}

Shortly after the discovery of PSR J1906+0746, some hints of profile variations were already interpreted as first signs of relativistic spin-precession occuring. Using observations from the Nan\c cay, Arecibo and Green Bank Radio Observatories, we report here the measurement of pulse profile and polarimetric variations. Using the Rotating Vector Model, we show that PSR J1906+0746 is likely to be an orthogonal rotator ($\alpha \simeq 80^\circ$). Fitting our polarimetric data to a precession model, we determined the geometry of the pulsar and found a wide misalignment angle ($\delta = 89_{-44}^{+85}$\,deg, 95\% C.L.), although the uncertainty is large. Assuming this geometry, we constructed the beam maps of both magnetic poles.  

%Shortly after the discovery of PSR J1906+0746, it was noted that it could suffer from geodetic precession. Indeed, the first measurements shows profile and polarimetric variations. Modeling our polarimetric data to the precession model by \cite{K09}, we can derive the geometry of the pulsar arguing for an orthogonal rotator. The misalignment angle is also large. Finally we constructed the beam maps of both pulsar's magnetic poles.
 
\end{abstract}

% add below any authors, subjects and objects for indexing 
%   add more lines if necessary
%   but leave all lines commented out
%\index[author]{LastName1, Initials|textbf}
%\index[author]{LastName2, Initials|textbf}
%\index[subject]{Keyword1}
%\index[subject]{Keyword2}
%\index[object]{Object1}
%\index[object]{Object2}

\firstsection % if your document starts with a section,

\section{Introduction}
PSR J1906+0746 is a young pulsar in a 4-hr orbit around a massive companion \cite{L06}. With a pulsar mass $M_{p}=1.323\pm0.011$\,M$_{\odot}$ and a companion mass $M_{c}=1.290\pm0.011$\,M$_{\odot}$  derived from timing measurement by \cite{K12}, one can estimate the relativistic spin-precession period to be 165 years assuming General Relativity.

Relativistic spin-precession in binary pulsars is a long-known effect
that is due to spin-orbit coupling 
(\cite[Damour \& Ruffini 1974; Barker \& O'Connell 1975]{D74,B75}).
The consequence of this precession is that our line of sight crosses different parts of the radio beam with time. Hence we can expect pulse shape and polarization variations.

%a sweep of the radio beam over our line of sight.
% By comparison, the geodetic precession period of the double pulsar is about 75 years REF %\cite{}%.

The non-detection of the interpulse
in archival data %from 1998 
and an increase of the signal-to-noise ratio (SNR) for the main
pulse between 1998 and 2005, suggested the first sign of change in the 
beam orientation with respect to our line of sight \cite{L06}.
Profile shape variations were also reported by \cite{K08} when they presented their timing solution.
More recently, they reported a preliminary beam map for the main pulse (Kasian 2012)\nocite{K12}.

We present in these proceedings further observations that allow us to determine the geometry of the pulsar and produce improved maps of its radio beam.
%So far, relativistic spin-precession has enabled us to map the beam shape of a handle of pulsars: 

\section{Observations}

This pulsar was observed from 2005 to 2009 with the BON backend at the
Nan\c cay Radio Telescope 
(hereafter NRT; \cite[Desvignes 2009]{D09})  and  the WAPPs, ASP and  GASP backends at the  Arecibo 305-m Observatory and  the Green Bank Telescope respectively (for further details, see Kasian 2012\nocite{K12}). All these observations were done at L-Band.
 Only the NRT observations provide calibrated  full Stokes profiles that were de-Faradayed with a Rotation Measure of 149 rad m$^{-2}$ and hence were used for the polarimetric study.
Given the low SNR of the NRT daily observations, the NRT profiles were integrated to form 13 profiles spanning 3 to 6 months to obtain reliable polarimetric fits.
% Before summing the archives in frequency, the weighted mean Rotation Measure was calculated to 149 rad m$^{-2}$ and applied to the data.
%Before summing the data in frequency, they were defaraded using the 

\begin{figure}
  \begin{center}

  \includegraphics[height=9cm, angle=-90]{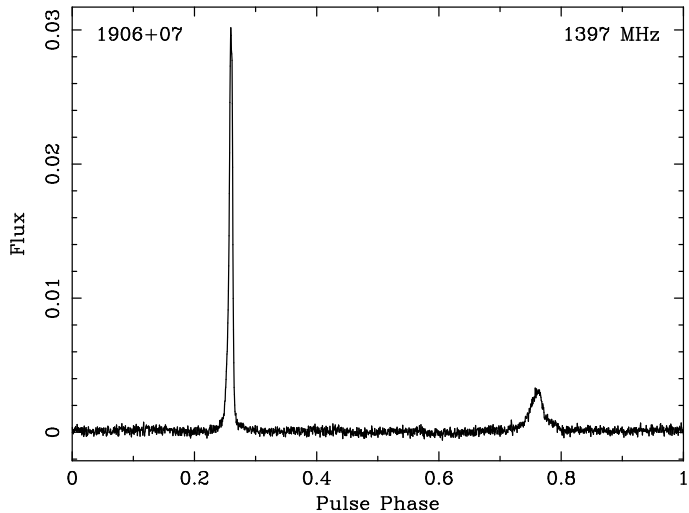}
  
  \caption{Mean pulse profile of PSR J1906+0746 as recorded in 2005 with the NRT.}
  \label{template}
  \end{center}
\end{figure}

\section{Profiles and polarimetric changes}

The mean pulse profile at the beginning of our dataset in 2005 consists of 2 sharp pulses separated by almost $180^\circ$ (see Fig. \ref{template}).
Over the 3 years course of the data span, we measured a change in the separation of the two pulses to be $2.1^\circ \pm 0.1^\circ ~\mathrm{yr}^{-1}$.

The flux density of both components was also estimated for each dataset using the radiometer equation, e.g. \cite{L05}.
It  decreased by a factor of $\sim $3 and $\sim $4.5 for the main pulse and the interpulse respectively. 

%They all agree. The Arecibo, GBT, Nan\c cay for the main pulse. 3... 4.5
%Also the flux density of the interpulse decrease by a factor XXX making it hard to detect with the Nan\c cay dataset.

Our polarization data first confirmed the high degree of linear polarization noticed in the discovery paper \cite{L06}. However the circular polarization under the main pulse gradually vanished between our first and last epochs.
According to the Rotating Vector Model (RVM) put forward by \cite{R69}, the typical 'S' curve of the Polarization Position Angle (PPA) can be described in terms of the geometry of the pulsar:

\begin{equation}
\label{eq}
\tan(\psi-\psi_{0})=\frac{\sin \alpha \sin( \phi-\phi_{0})}{\sin( \alpha+\beta) \cos \alpha-\cos(\alpha+\beta)\sin \alpha\cos(\phi-\phi_{0})},
\end{equation}
where $\alpha$ is the angle between the rotation and magnetic axis and $\beta$ denotes the impact parameter. Here $\psi$ is the measured PPA at the longitude $\phi$, $\phi_{0}$ the longitude under the magnetic axis at the closest approach of the line of sight and $\psi_{0}$ the PPA at the longitude $\phi_{0}$.

% Method 1
Fitting the RVM to each of our 13 polarimetric profiles, we measured  $\alpha$ to be close to $80^\circ$ for all epochs. A constant value of $\alpha$ is expected and this result strongly suggests an orthogonal rotator, with the two pulses representing the cone of emission of both magnetic poles.
The RVM results also show a small increase of $\beta$ with time (i.e. the slope of the PPA under the main pulse in decreasing with time),  indicating that our line of sight is moving away from the magnetic poles.
More importantly, a marginal decrease of $\psi_{0}$ is also detected.

In the next section, we used this change in $\psi_{0}$ to determine the geometry of the system and map the radio emission beams.

\begin{figure}
  \begin{center}
    \includegraphics[height=10cm, angle=-90]{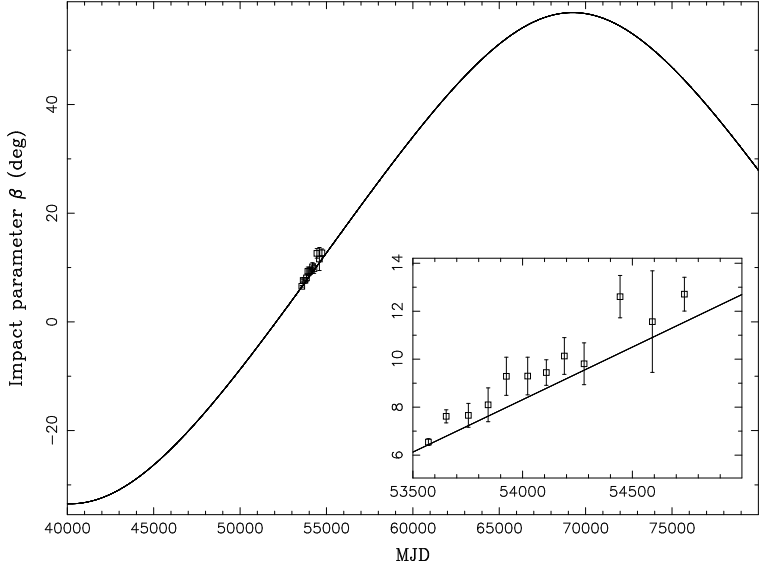}
    \caption{Impact parameter $\beta$ as a function of MJD. The black boxes show the impact parameter as determined by the simple RVM fits. The black line represents the model from the best fit values given by Table \ref{tab1}. The inset shows a zoom over the data.}
  \label{beta-model}
  \end{center}
\end{figure}

\section{Modeling of relativistic spin precession}

% Method 2
%Absolute value of PPA $\psi$  should change with time.
%the method described in details in \ref{K09}. The results of this global model are given in Table \ref{tab1}.

 \cite{K09} have shown that the absolute value of the  PPA $\psi_{0}$  should change with time. Applying their global precession model and fitting for the magnetic inclination angle $\alpha$, the misalignment angle  $\delta$, the reference precessional phase $\Phi_{SO}^{0}$  plus the 13 phase offsets give the results reported in Table \ref{tab1}.
 
\begin{table}[b]
  \caption{Results of the global precession model at a 95\% confidence level.}
  \label{tab1}
  \begin{center}
  \begin{tabular}{c c c c}
    \hline
    $\alpha$ & $\delta$  & $\Phi_{SO}^{0}$ & $\chi_{red}^{2}$ \\
    \hline
    $81_{-66}^{+1}$ & $89_{-44}^{+85}$ & $ 83_{-93}^{+117}$ & $2.22$\\
    \hline
  \end{tabular}
  \end{center}
\end{table}

The magnetic inclination angle is consistent with the value determined with the simple RVM fit. The large value of the misalignment angle explains our quick detection of the precession effects. Its large uncertainty can be justified by the fact that the impact parameter did not have a sign reversal as it happened for PSR J1141$-$6545 \cite{M10}. 

With the geometry of the system derived, we can now  produce a map of the emission beam.
First, a set of gaussians is fitted to all profiles. The height of the gaussians are normalized using flux density measurements.

 When producing the beam map for PSR J1141$-$6545, Manchester \etal\, (2010) \nocite{M10} aligned the pulse profiles using the edges.
In this work, the  profiles are aligned with respect to the magnetic poles based on the individual measurements of $\phi _{0}$ given by the simple RVM fits, hence making no assumption on the beam shape.
This alignment based on polarimetric results explains the offset in the main pulse longitude between this beam map and the one produced by \cite{K12}.

\begin{figure}
  \begin{center}

    \begin{minipage}{0.5\textwidth}
     \begin{center}
      $\vcenter{\hbox{\includegraphics[height=\textwidth, angle=-90]{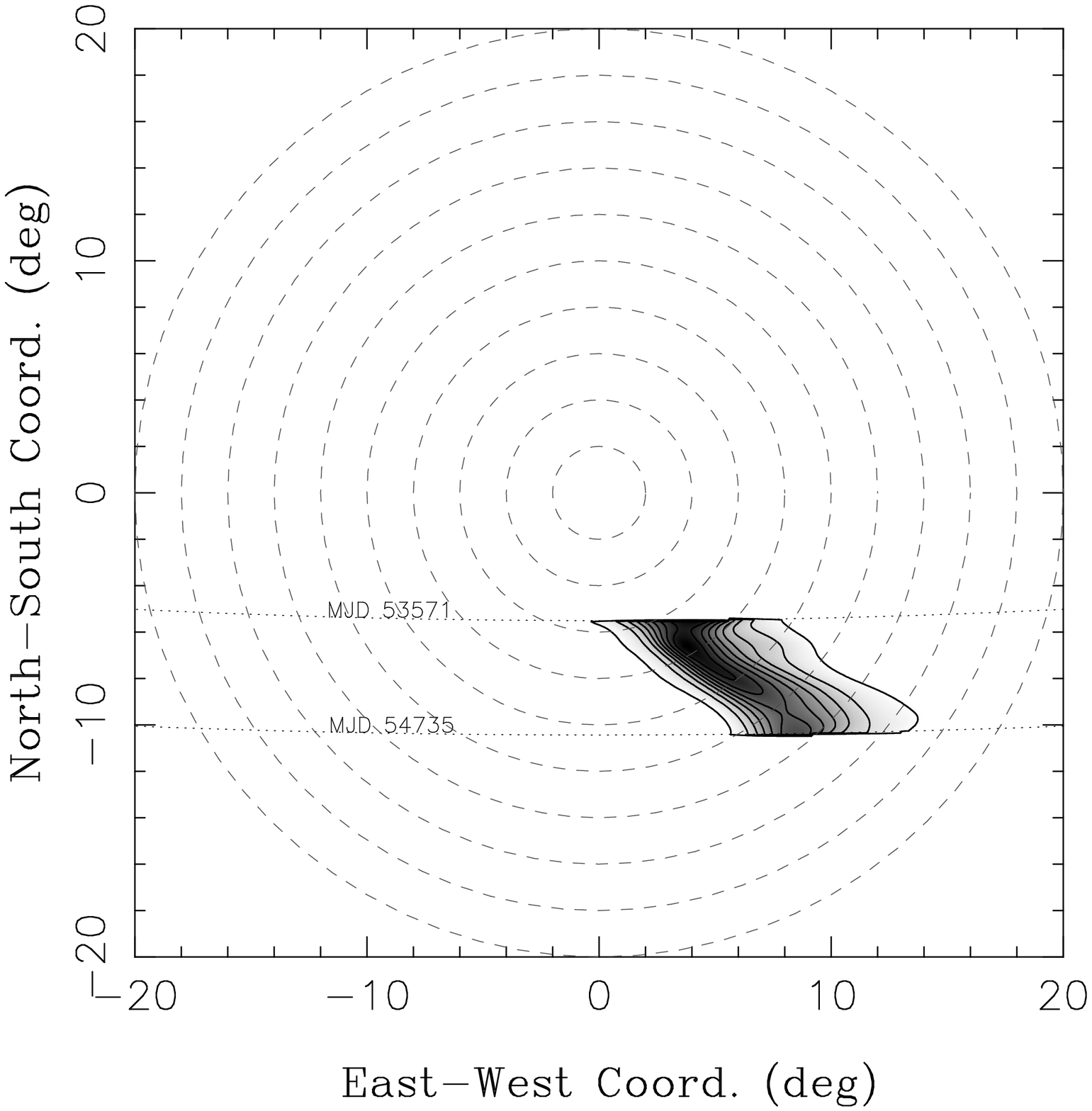}}}$
     \end{center}
    \end{minipage}%
    \begin{minipage}{0.5\textwidth}
     \begin{center}
      $\vcenter{\hbox{\includegraphics[height=\textwidth, angle=-90]{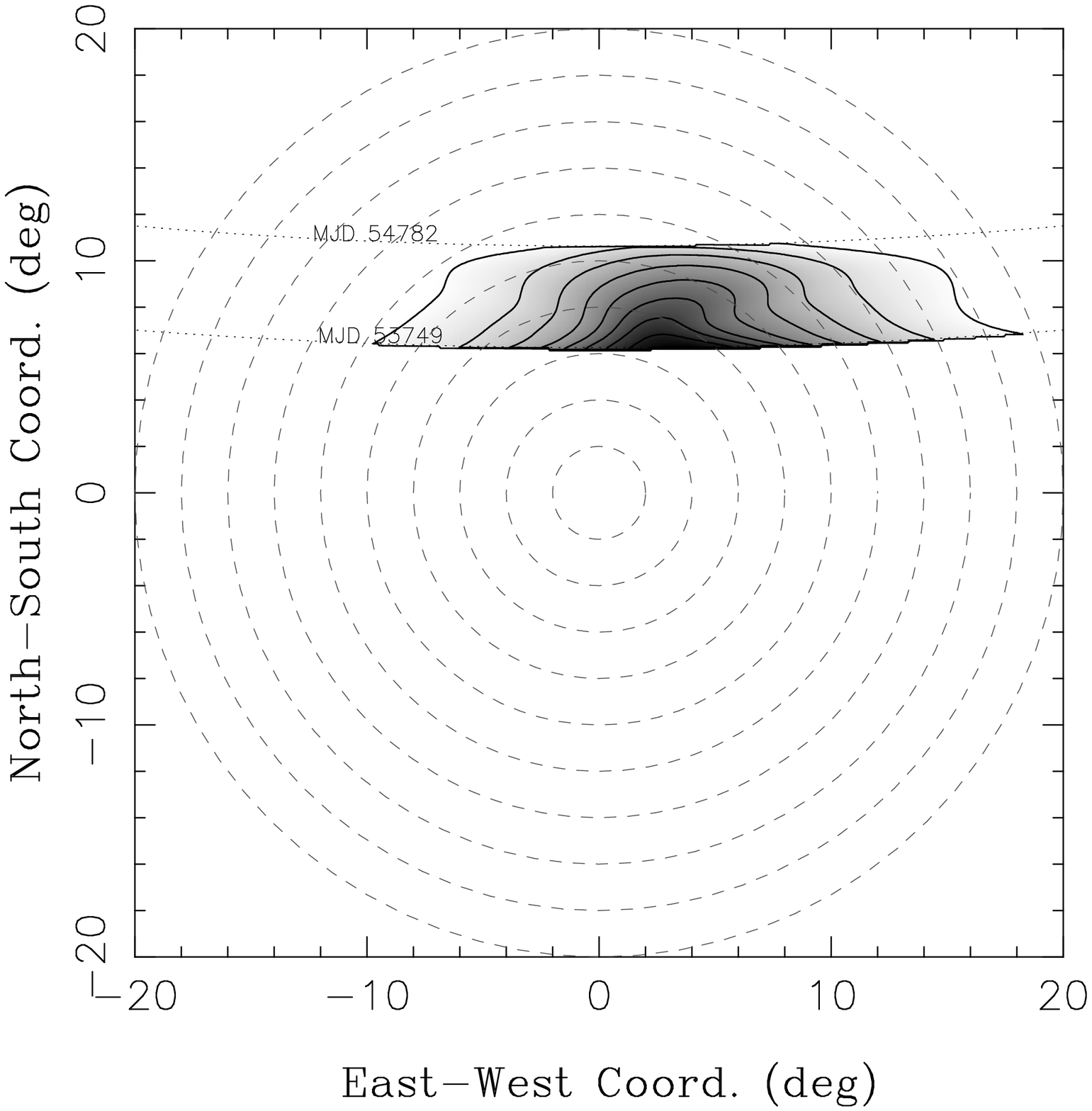}}}$
     \end{center}
    \end{minipage}
    
    \caption{{\it Left panel:} Beam map of the main pulse. The gray contour plot shows an axial emission with a decrease in intensity as the line of sight moves away from the magnetic pole.
    {\it Right panel:} Beam map of the interpulse. In both plots, the black lines represent the increments of emission, in steps of 10\%.}
    \label{beams}
  \end{center}
\end{figure}

The results of the beam maps are shown Fig. \ref{beams}.
In the case of the main pulse, we see axial emission with the flux decreasing as the line of sight is moving away from the magnetic pole. For the interpulse, the emission is more extended.

The beam maps clearly show the change in the separation of the two components as the line of sight is moving away from the magnetic poles.
These profile variations will undoubtedly have an impact on the timing study of this pulsar (van Leeuwen \etal{,} in prep.) beyond the usual timing noise in young pulsars \cite{L10}.

\begin{thebibliography}{}


\bibitem[Barker \& O'Connell (1975)] {B75}
{{Barker}, B.~M. and {O'Connell}, R.~F.} 1975,
\textit{ApJ}, 199, L25

\bibitem[Damour \& Ruffini (1974)]{D74}
{{Damour}, T. and {Ruffini}, R.} 1974,
\textit{Academie des Sciences Paris Comptes Rendus Serie Sciences Mathematiques}, 279, 971-973

\bibitem[(Desvignes 2009)]{D09}
{Desvignes, G.} 2009,
\textit{Th\`ese de doctorat, Universit\'e d'Orl\'eans}

\bibitem[Kasian (2008)]{K08}
{Kasian, L.} 2008,
\textit{American Institute of Physics Conference Series}, 983

\bibitem[Kasian (2012)]{K12}
{Kasian, L.} 2012,
\textit{PhD thesis, University of British Columbia}

\bibitem[Kramer \& Wex (2009)]{K09}
{{Kramer}, M. and {Wex}, N.} 2009,
\textit{Classical and Quantum Gravity}, 26, 073001

\bibitem[Lorimer \& Kramer (2005)]{L05}
{Lorimer, D. and {Kramer}, M.} 2005,
\textit{Handbook of Pulsar Astronomy}

\bibitem[(Lorimer \etal\ 2006)]{L06}
{{Lorimer}, D.~R. and {Stairs}, I.~H. and {Freire}, P.~C.~C. et al. } 2006,
\textit{ApJ}, 640, 428-434

\bibitem[(Lyne \etal\ 2010)]{L10}
{{Lyne}, A. and {Hobbs}, G. and {Kramer} et al.} 2010,
\textit{Science}, 329, 408-

\bibitem[(Manchester \etal\ 2010)]{M10}
{Manchester, R.~N., Kramer, M., Stairs, I.~H.  et al.} 2010,
\textit{ApJ}, 710, 1694-1709

\bibitem[Radhakrishnan \& Cooke (1969)]{R69}
{Radhakrishnan, V. and Cooke, D.~J. } 1969,
\textit{ApL}, 3, 225-229

\end{thebibliography}
\end{document}